\documentclass[pra,showpacs,aps,twocolumn]{revtex4}

\usepackage{psfrag,graphicx}
\usepackage{amsmath,amssymb}

\begin{document}
\title{Grover like Operator Using Only Single-Qubit Gates}
\author{G.~Kato}
\email{kato@theory.brl.ntt.co.jp}
\affiliation{NTT Communication Science Laboratories,\\
NTT Corporation \\
3-1, Morinosato Wakamiya, Atsugi-shi, Kanagawa Pref., 243-0198 Japan
}
\date{\today}

\begin{abstract}
We propose a new quantum circuit for the quantum search problem. The
 quantum circuit is
superior to Grover's algorithm in some realistic cases. 
The reasons for
the superiority are in short as follows: 
In the quantum circuit proposed in this paper, all the operators except
for the oracle can be written as direct products of single-qubit gates.
Such separable operators
can be executed much faster than multi-particle operators, such as
 c-NOT gates and 
 Toffoli gates, in many realistic
systems.  The
idea of this quantum circuit is inspired by the Hamiltonian used in the adiabatic quantum computer.
In addition, the scaling of the number of oracle calls for this circuit
is the same as that for Grover's algorithm, i.e. $O\left(2^{n/2}\right)$.
\end{abstract}

\maketitle

\section{introduction}
Since the concept of the quantum computer (QC) was
proposed \cite{B82,D85,F85}, many quantum
algorithms \cite{DJ92,S94,G97,TKM05} that are superior to classical
algorithms have been proposed.  These algorithms have inspired 
many researchers, and the number of the researchers 
investigating the QC has increased dramatically as a result.

Though many results are generated daily, there remains a serious
problem. The generated quantum circuits utilize
the properties of quantum mechanics effectively, but almost all of then are
modifications or combinations of just three
quantum circuits based on quantum Fourier transformation \cite{K95}, 
quantum amplitude amplification \cite{BHM+00} or discrete quantum
random walk \cite{AAJ+01}. This indicates that it is very hard to
design new quantum circuits that use the properties of quantum mechanics
effectively.

Recently, some frameworks differing from the QC have been proposed,
 such as the adiabatic quantum computer (AQC)\cite{FGG+00}, and the
continuous random walk \cite{FG97}, and many results have been
 forthcoming in
this area.  In this paper, we focus on the AQC, whose procedure is
identified by a Hamiltonian.  Recently, it was proved that the
 calculation power of  the AQC has
the same as that of the QC \cite{ADK+04}. This means that the QC can be
emulated using the AQC and vice versa with polynomial time and space
with respect to the input size.  On the other hand, the properties of
the problems that the QC and AQC are good at are different.  These two facts indicate that new concepts of quantum circuits
must be given by the explicit modification from the Hamiltonians for the AQC
into finite size quantum circuits for the QC. We think this is a good
strategy for designing new quantum circuits that use the properties of quantum
mechanics effectively.

In this paper, we propose a new quantum circuit modified from a Hamiltonian
for the AQC. This is the first simple example of a quantum circuit
obtained by following the
above strategy.  Here, we treat the well-investigated problem in
the QC, i.e., the quantum search problem, in order to check the
efficiency of the strategy.  As a result, we get a new quantum circuit
that is
superior to the quantum circuit used in Grover's algorithm in some
cases. 
The Hamiltonian just gives us 
some hints, and the new quantum circuit is intuitively generated using those
hints. Thus, we can not show some explicit procedures for the
modification.

Here, we have to mention that, from the past work \cite{L96}, quantum circuits for
the QC can be easily modified from the Hamiltonians for the AQC, but the
quantum circuits generated by the modification simply follow the time evolution of
the AQC. Consequently, such quantum circuits are very redundant and inefficient for
realistic calculations. The quantum circuits that we want to modify from
the Hamiltonians are not such useless quantum circuits but practical ones.

To avoid any confusion, we should clarify that our circuit is superior in that 
it may be executed faster than Grover's
algorithm in realistic systems since it uses only simple
operators, each of which rotate just one-qubit, except
for the oracle. However, the circuit does not offer reduced
complexity. 
Actually, both it and
Grover's algorithm have exactly the same complexity $O\left(2^{n/2}\right)$. 
For these reasons, the superiority of the new circuit will be meaningful mainly to experimentalists. 

In
Sec. \ref{sec:Grover_algorithm}, we briefly review Grover's algorithm to
facilitate  comparison between it and the
expressions for the new quantum circuit.
 In Sec. \ref{sec:G_algorithm}, we show the explicit form of
the new quantum circuit and prove that the quantum circuit can execute quantum search
efficiently.  In Sec. \ref{sec:numerical_calculation}, we numerically
simulate the new quantum circuit to show how well it executes
quantum search.  In Sec. \ref{sec:relation_AQC}, we show the relations
between quantum circuits and Hamiltonians for the AQC. These relations are the hints
for generating the new quantum circuit. The last section summarizes
our conclusions. Technical details of a proof are in Appendix
\ref{sec:proof_of_lemma}.

\section{Grover's algorithm}
\label{sec:Grover_algorithm}
By Grover's algorithm, the quantum search problem can be solved.
This means that we can find integer $j$ from $0$ to $2^n-1$ using the
oracle operator $\hat {O_r}$ such that
\begin{eqnarray}
\hat{O_r}\left|m\right>\otimes\left|k\right>
&:=&
\left|m\right>\otimes\left|k\oplus\delta\left(m,j\right)\right>
\label{eq:def_oracle}
\end{eqnarray}
by the algorithm.
The operator $\hat {O_r}$ acts on two registers: one is
$2^n$-dimensional, corresponding to the search space, and the other is
$2$-dimensional, corresponding to the output of the oracle.
 Grover's algorithm can be expressed as follows.
First, we generate the initial state
\begin{eqnarray}
\left|\bar0\right>&:=&2^{-\frac n2}\sum_{m=0}^{2^n-1}\left|m\right>.
\label{def:initial_state}
\end{eqnarray}
Next,
we iterate the
 two operations, which are identified by the following operator:
\begin{eqnarray}
\hat G&:=&1-2\left|\bar 0\right>\left<\bar 0\right|,\\
\hat O&:=&1-2\left|     j\right>\left<     j\right|.
\label{eq:def_oracle_Gr}
\end{eqnarray}
Note that, in general $\hat G\!\cdot\!\hat O$ is written by $G$, e.g., \cite{NC00}, and is
called the Grover operator. The number of iterations is
\begin{eqnarray}
N&:=&\left[\frac{\pi}{4\arcsin 2^{-\frac n2}}\right],
\end{eqnarray}
where $[r]$ indicates the integer part of real number $r$.  Note that the
operator $\hat O$ (\ref{eq:def_oracle_Gr}) is outwardly different from
the oracle operator $\hat{O_r}$ (\ref{eq:def_oracle});
however, $\hat O$ can be simulated from one use of $\hat{O_r}$
by using the second register as an ancilla prepared in state $\frac1{\sqrt2}\left|0\right>-\frac1{\sqrt2}\left|1\right>$.
 Finally, we observe the state
using the computational basis, i.e., $\left|0\right>$,
$\left|1\right>$,$\cdots$ ,$\left|2^n-1\right>$.  The success
probability of Grover's algorithm, i.e., the probability to detect the
state $\left|j\right>$, goes to $1$ in the limit 
$n\rightarrow\infty$. This is equivalent to the following relation:
\begin{eqnarray}
\lim_{n\rightarrow\infty}
\left|\left<j\right|\left(\hat G\!\cdot\! \hat O\right)^N\left|\bar 0\right>\right|^2
&=&1.
\label{eq:math_glover_algirism}
\end{eqnarray}
The scaling of the success
probability versus $n$ is
$1-O\left(2^{-n}\right)$.
The relation (\ref{eq:math_glover_algirism}) can be easily proved as follows.

{\it Proof}:

The operator $\hat G\!\cdot\!\hat O$ modifies any vector in the space
spanned by $\left|\bar 0\right>$ and $\left|j\right>$ into another
vector in the same space. Then, we restrict the Hilbert space to the
two dimensional space, i.e.,
$\left\{\left|\psi\right>=a\left|\bar0\right>+b\left|j\right>\right\}$
in this proof.  Under this restriction, the operator
$\hat G\!\cdot\!\hat O$ can be written as the following two dimensional
matrix:
\begin{eqnarray}
\hat G\!\cdot\!\hat O 
&=&
-
\begin{pmatrix}
 1-2^{1-      n }               & -2^{1-\frac n2}\sqrt{1-2^{-n}} \\
   2^{1-\frac n2}\sqrt{1-2^{-n}}&1-2^{1-      n } 
\end{pmatrix}
\nonumber\\
&=&
-
\begin{pmatrix}
 \cos2\theta&-\sin2\theta \\
 \sin2\theta& \cos2\theta
\end{pmatrix},
\\
\theta&:=&\arcsin2^{-\frac n2}\quad\quad 0<\theta\leq\frac\pi2.
\label{def:theta}
\end{eqnarray}
Here, we use the basis
$\left\{\left|\bar 0\right>,\left|\bar1\right>:=\frac{\left|j\right>-2^{-\frac n2}\left|\bar 0\right>}{\sqrt{1-2^{-n}}}\right\}$.
From this expression, it is easy to show that
\begin{eqnarray}
&&\left|\left<j\right|\left(\hat G\!\cdot\!\hat O\right)^N\left|\bar 0\right>\right|^2
\nonumber\\
&=&
\sin^2 \left(2N+1\right)\theta
\nonumber\\
&=&
\sin^2\left(2\left[\frac{\pi}{4\arcsin 2^{-\frac n2}}\right]+1\right)
\arcsin2^{-\frac n2}.
\end{eqnarray}
From the last equation, it is clear that relation
(\ref{eq:math_glover_algirism}) holds. $\square$

\section{Quantum search algorithm  using a new quantum circuit}
\label{sec:G_algorithm}
\subsection{The case of one solution}
We propose a new quantum circuit by which the Grover iteration can be replaced.

The outline of the algorithm is the same as Grover's algorithm, but
in order to avoid misunderstanding we show whole algorithm below.
First, we prepare the initial state $\left|\bar0\right>$, which is the
same as the initial state of Grover's algorithm. Next, we iterate the
 two operations, which are identified by the following operator:
\begin{eqnarray}
\hat G'
&:=&
\exp\left(\varphi\left(\omega\right)
\sum_{\alpha=0}^{n-1}S_x^{(\alpha)} i\right),
\label{eq:def_oracle_Go_G'}\\
\hat O'
&:=&
\exp\left(\omega \left|j\right>\left<j\right| i\right).
\label{eq:def_oracle_Go}
\end{eqnarray} 
The number of iterations is
\begin{eqnarray}
 N'&:= &\left[\frac{\pi}{4 \sin\left|\frac\omega2\right|}2^{\frac n2}+\frac12\right].
\label{eq:suggestion_2} 
\end{eqnarray}
The variable $\omega$ in the above definition can be chosen from the
region $-\pi<\omega<\pi$ and is independent of $n$ and $j$.  The operator
$S_x^{(\alpha)}$ and the function $\varphi\left(\omega\right)$ are
defined
later.  Finally, we observe the state using the computational
basis. The success probability of this algorithm goes to $1$ in the
limit $n\rightarrow \infty$.  This is equivalent to the following
relation:
\begin{eqnarray}
\lim_{n\rightarrow\infty}
\left|\left<j\right|\left(\hat G'\!\cdot\!\hat O'\right)^{N'}\left|\bar 0\right>\right|^2
&=&1.
\label{eq:suggestion_1}
\end{eqnarray}
The scaling of the success
probability versus $n$ is
$1-O\left(n^{-1}\right).$
A proof of relation (\ref{eq:suggestion_1}) is located at the end of
 this section. 

 Here,
we have to note three things.  First,
the scaling of the number of oracle calls is $O\left(2^{n/2}\right)$
for any $-\pi<\omega<\pi$ when $\omega\neq0$.
Here, we have to point out that
the operator $\hat O'$ (\ref{eq:def_oracle_Go}) can actually be simulated by a constant number of
calls to the oracle $\hat{O_r}$ (\ref{eq:def_oracle}), where the number depends on $\omega$.
A method of simulation is as follows.
We introduce
a naturally generalized oracle as
\begin{eqnarray}
 \hat {O_r}'\left|m\right>\otimes\left|k\right>
&:=&\left|m\right>\otimes\left|k+\delta\left(m,j\right)\makebox{ mod }\omega_d\right>
\label{eq:def_oracle_ge}
\end{eqnarray}
for arbitrary integer $\omega_d$. 
The operator $\hat {O_r}'$ (\ref{eq:def_oracle_ge}) acts on two registers:
 one is $2^n$-dimensional and the other is
$\omega_d$-dimensional.
It is easy to show that this operator $\hat {O_r}'$
 can be simulated by a constant number of
calls to the oracle $\hat{O_r}$. Furthermore,
the operator $\hat O'$ can be simulated from one use of $\hat{O_r}'$ 
by using the second register as ancillae prepared in state
\begin{eqnarray}
\sum_{k=0}^{\omega_d-1}\exp\left(\frac{k\omega_c}{\omega_d}2\pi i\right) \left|k\right>.
\end{eqnarray}
In this definition, $\omega_d$ and $\omega_c$ are chosen so as to satisfy
$\frac{\omega_c}{\omega_d}2\pi=\omega$.
Then, the operator $\hat O'$ can be simulated by $\hat{O_r}$.
Second,
 the difference in execution time between $\hat {O_r}$
 and $\hat O'$ 
 probably 
will not
depend on $n$  in most cases.
This expectation comes form the following consideration.
Once we know the explicit
 circuit for $\hat {O_r}$, we will probably be able to make a circuit
 corresponding to $\hat {O_r}'$ in such a way that the difference of
 execution time of these two circuits 
does not
depend on $n$.
This expectation has no meaning from a computer science point of view,
 since the oracle $\hat{O_r}$ is usually treated as a black-box.
However, in case of actual calculations using a real system, it is important to
 think in term of the execution time of operations.
Third, 
when $\omega=\pm\pi$, the relation (\ref{eq:suggestion_1}) does not hold.
This is related to the fact that the value $\omega$ influences not
only the number of iterations $N'$ but also the speed of
the convergence (\ref{eq:suggestion_1}). 
  For example, when $\omega$ approaches $\pm\pi$,
 the speed of the convergence decreases.
 On the other hand, when $\omega$ approaches $0$, 
 the speed of the convergence increases. 
Here, the change in the speed means the change in the constant factor of  the scaling. 

Here, we define the function $\varphi\left(\omega\right)$ and the operator $S_x^{(\alpha)}$ used in
the above outline of the algorithm. First, $S_x^{(\alpha)}$ is the
operator which acts only on the $\alpha$-th qubit, and the action on the qubit
can be written as $\begin{pmatrix}0&\frac12 \\\frac12&0\end{pmatrix}$ 
using the computational basis. Therefore, we can write
$S_x^{(\alpha)}$ as follows:
\begin{eqnarray}
S_x^{(\alpha)}&:=&Id\otimes \cdots\otimes 
\begin{pmatrix}
0&\frac12 \\\frac12&0
\end{pmatrix}
\otimes\cdots\otimes Id.
\end{eqnarray}
Note that $2S_x^{(\alpha)}$ is simply Pauli operator $\sigma_x$ applied on qubit $\alpha$.
Next, we define $\varphi\left(\omega\right)$ implicitly as follows:
\begin{eqnarray}
\cot\frac\omega2&=&\sum_{s=1}^{n}P_n\left(s\right)\cot\frac{s\varphi\left(\omega\right)}2,
\nonumber
\end{eqnarray}
\vspace{-.5cm}
\begin{equation}
\makebox[1cm]{}-\frac{2\pi}{n}<\varphi\left(\omega\right)<\frac{2\pi}{n},\quad\operatorname{sgn}\left(\omega\right)=\operatorname{sgn}\left(\varphi\left(\omega\right)\right)
\label{eq:definition_alpha_2}
\end{equation}
where
\begin{eqnarray}
P_n\left(s\right)&:=&\frac{n!2^{-n}}{s!\left(n-s\right)!}. 
\end{eqnarray}
Recall that $2^n$ is the number of elements in the set from which item
$j$ is selected and that $\omega$ is an arbitrary number in the region
$-\pi<\omega<\pi$. Note that the function $\varphi\left(\omega\right)$
depends on $n$.
\begin{figure*}
\psfrag{labelx}{$\omega$}
\psfrag{labely}{$\varphi\left(\omega\right)$}
\psfrag{ 0.2pi}{$ 0.2\pi$}
\psfrag{ 0.1pi}{$ 0.1\pi$}
\psfrag{-0.1pi}{$-0.1\pi$}
\psfrag{-0.2pi}{$-0.2\pi$}
\psfrag{2pi}{$ 2\pi$}
\psfrag{1pi}{$ \pi$}
\psfrag{zero}{$ 0$}
\psfrag{-1pi}{$-\pi$}
\psfrag{-2pi}{$-2\pi$}
\includegraphics[scale=0.9]{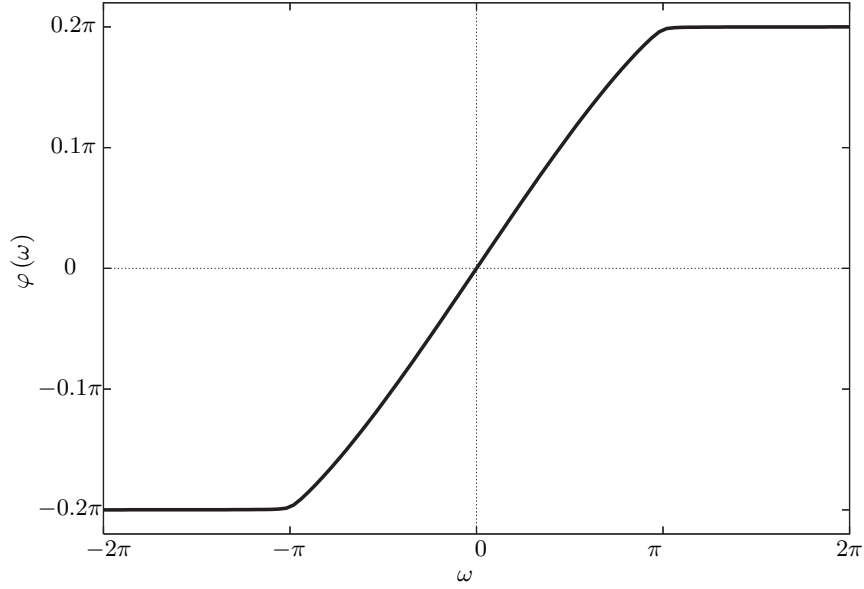}
\caption{\label{fig:varphi}
A plot of the function $\varphi\left(\omega\right)$ defined by
 (\ref{eq:definition_alpha_2}) for $n=10$
}
\end{figure*}
As an example, a plot of the function $\varphi\left(\omega\right)$ for $n=10$
is shown in Fig \ref{fig:varphi}.

The rest of this section is devoted to proving relation
(\ref{eq:suggestion_1}).

{\it Proof:}

First of all, we show the main idea underlying this proof in order to
provide
 some
insight into why it works. The idea consists of three parts.
First, $\hat G'\hat O'$ leaves $\tilde S^2$ (\ref{eq:modified_total_spin})
eigenspaces invariant, and both $\left|\bar 0\right>$ and
$\left|j\right>$ lie in the same eigenspace, so we can restrict our
study to this eigenspace.
Second, $\left|\bar 0\right>$ and $\left|j\right>$ have most of their
support on the $2$-dimensional subspace spanned by two particular
eigenstates of $\hat G'\hat O'$, $\left|\psi_{\gamma_\pm}\right>$  whose
eigenvalues are $\gamma_\pm$ (\ref{eq:inportant_eignevalue}),
so we can even more restrict our study to this subspace.
 Finally, due to the corresponding eigenvalues
$\gamma_\pm$, we need to repeat $\hat G'\hat O'$ a certain number of
times (\ref{eq:suggestion_2}) to rotate
$\left|\bar 0\right>$ to
$\left|j\right>$.
Based on this idea,
we obtain a strict proof as follows.

The operator $\hat G'\!\cdot\!\hat O'$ is a block diagonal matrix in the
case of
the computational basis and each block can be characterised by eigenvalues
of the operator
\begin{eqnarray}
\tilde S^2:= 
       \left(\sum_{\alpha=0}^{n-1}       S_x^{(\alpha)} \right)^2
\!\!+\!\left(\sum_{\alpha=0}^{n-1}\tilde S_y^{(\alpha)} \right)^2
\!\!+\!\left(\sum_{\alpha=0}^{n-1}\tilde S_z^{(\alpha)} \right)^2\!\!\!,
\label{eq:modified_total_spin}
\end{eqnarray}
where
\begin{eqnarray}
&&{}\!\!\!\!\!\!\!\!\!\!\! 
\tilde S_y^{(\alpha)}:=
\left(-\right)^{j^{(\alpha)}}\!\!\!\!
Id\otimes \cdots\otimes 
\begin{pmatrix}
0&-\frac12 i \\\frac12 i&0
\end{pmatrix}
 \otimes\cdots\otimes Id,\nonumber\\
&&{}\!\!\!\!\!\!\! \!\!\!\!
\tilde S_z^{(\alpha)}:=
\left(-\right)^{j^{(\alpha)}}\!\!\!\!
Id\otimes \cdots\otimes 
\begin{pmatrix}
\frac12&0 \\0&-\frac12
\end{pmatrix}
 \otimes\cdots\otimes Id
\label{eq:modified_Pauli}
\end{eqnarray}
and $j^{(\alpha)}$ is $0$ or $1$ such that
\begin{eqnarray}
 j&=&\sum_{\alpha=0}^{n-1}2^\alpha j^{(\alpha)}.
\end{eqnarray}
Note that the operators $2\tilde S_y^{(\alpha)}$ and $2\tilde S_z^{(\alpha)}$ defined by (\ref{eq:modified_Pauli}) reduce to the Pauli operators in the
special case j=0. Otherwise, the operators $2\tilde S_y^{(\alpha)}$
and $2\tilde S_z^{(\alpha)}$ are equivalent to the Pauli operators up to an overall
phase.
The states $\left|\bar 0\right>$ and $\left|j\right>$ belong to the
subspace whose eigenvalue for $\tilde S^2$ is
$n\left(n+2\right)/4$.
This subspace reduces to the maximal total spin subspace in the special case $j=0$.
  In the rest of this section, we
restrict the Hilbert space to this subspace and use the following two
bases
\begin{eqnarray}
\left\{
 \left|s_x\right>
\left|\sum_{\alpha=1}^n       S_x^{(\alpha)}\left|s_x\right>=\left(-s+n/2\right)\left|s_x\right>
\right.
\right\},\\
\left\{
 \left|s_z\right>
\left|\sum_{\alpha=1}^n\tilde S_z^{(\alpha)}\left|s_z\right>=\left(-s+n/2\right)\left|s_z\right>
\right.
\right\}.
\end{eqnarray}
Note that it is easy to see that
$\left|\bar0\right>\propto\left|0_x\right>$ and
$\left|j\right>\propto\left|0_z\right>$. Then, over all phases are
defined in such a way that $\left<s_x|0_z\right>>0$,
$\left<0_x|s_z\right>>0$, $\left|\bar0\right>=\left|0_x\right>$ and
$\left|j\right>=\left|0_z\right>$.

The eigenvalues $ \exp\left(\gamma+\frac{n\varphi(\omega)}2\right) i$
and the corresponding eigenvectors $\left|\psi_\gamma\right>$ for
$\hat G'\!\cdot\!\hat O'$ satisfy the relation
\begin{eqnarray}
 \frac{\left<s_x|\psi_\gamma\right>}{1-\exp\left(\omega i\right)}
&=&\frac{\left<s_x|0_z\right>\left<0_z|\psi_\gamma\right>}{1- \exp
\left(\gamma+s\varphi\left(\omega\right) \right)i}.
\label{eq:original_relation}
\end{eqnarray}
Then, the following two relations hold:
\begin{eqnarray}
 \frac{1}{1-\exp\left(\omega i\right)}
&=&\sum_{s=0}^{n}
\frac{P_n\left(s\right)}
     {1- \exp\left(\gamma+s\varphi\left(\omega\right) \right)i},
\label{eq:definition_of_eigenvalue}\\
{}\!\!\!\!\!\!\!\!\frac{1}
     {
\left|1-\exp\left(\omega i\right)\right|^2
}
&=&\sum_{s=0}^{n}
\frac{P_n\left(s\right)\left|\left<0_z|\psi_\gamma\right>\right|^2}
     {
\left|1-\exp\left(\gamma+s\varphi\left(\omega\right) \right)i\right|^2
}.
\label{eq:definition_of_eigenvector}
\end{eqnarray}
In the derivation of the above two relations, we use the relation
\begin{eqnarray}
 \left|\left<s_x|0_z\right>\right|^2&=&P_n\left(s\right).
\end{eqnarray}

Next, we show that there are two eigenvalue series
$\exp\left(\gamma_\pm+\frac{n\varphi\left(\omega\right)}2\right)i$ for
the operator $\hat G'\!\cdot\!\hat O'$ such that
\begin{eqnarray}
 \lim_{n\rightarrow\infty}2^{\frac n2}\gamma_\pm&=&\pm2\sin\frac\omega2 
\label{eq:inportant_eignevalue},
\end{eqnarray}
where we regard $\gamma_\pm$ as two series with respect to $n$ defined
by $\omega$.  In order to prove this relation, we use the following
relation
\begin{eqnarray}
 \lim_{n\rightarrow \infty}\frac{n\varphi\left(\omega\right)}{2}&=&\omega.
\label{eq:asymptotics_of_alpha}
\end{eqnarray}
Recall that the function $\varphi\left(\omega\right)$ is defined by
(\ref{eq:definition_alpha_2}).  This relation is derived from
\begin{eqnarray}
 \lim_{n\rightarrow\infty}\sum_{s=1}^{n}P_n\left(s\right)\cot\frac{sr}{2n}
&=&
\cot\frac r4,
\label{eq:limit_of_definition_w_r_t_varphi}
\end{eqnarray}
where $-2\pi<r<2\pi$.  Relation
(\ref{eq:limit_of_definition_w_r_t_varphi}) is a special case of the
following Lemma.
\begin{itemize}
 \item {\it Lemma}:
\begin{eqnarray}
 \lim_{n\rightarrow\infty}\sum_{s=1}^{n}P_n\left(s\right)f\left(\frac{s}{n}\right)&=&f\left(\frac12\right),
\label{eq:limit_relation}
\end{eqnarray}
where $f\left(\zeta\in\mathbb{C}\right)$ is a meromorphic function in the region
$\left|\zeta-\frac12\right|<1/2+\delta$ and has only one pole at point
$\zeta=0$.
\end{itemize}

(A proof of this lemma is given in Appendix
\ref{sec:proof_of_lemma}.) Then, relation
(\ref{eq:asymptotics_of_alpha}) is proved.  Now, we define two functions
$g\left(n,\zeta\right)$ and $ g^{(q)}\left(\zeta\right)$
\begin{eqnarray}
g\left(n,\zeta\right)
&:=& 
\frac{1}{1-\exp\left(\omega i\right)}
\nonumber\\&&
-\sum_{s=0}^{n}
\frac{P_n\left(s\right)}
     {1- \exp\left(\zeta+s\varphi\left(\omega\right) \right)i},
\\
 g^{(q)}\left(\zeta\right)&:=&\frac1{q!}\frac {d^q}{d{\tilde\zeta}^q}
\left.
\frac{1}
     {1- \exp\left(\tilde\zeta+\zeta) \right)i}
\right|_{\tilde\zeta=0}.
\end{eqnarray}
It is clear that $g\left(n,\gamma\right)$ is equal to $0$ from
condition (\ref{eq:definition_of_eigenvalue}).
  Then, the sufficient condition of
(\ref{eq:inportant_eignevalue}),
\begin{eqnarray}
&&\lim_{n\rightarrow\infty}g\left(n,\zeta2^{-\frac n2}\right)2^{\frac n2}
\nonumber\\
 &=&
\frac1\zeta i-\lim_{n\rightarrow\infty}
\sum_{q=1}^\infty
\sum_{s=1}^{n}
 P_n\left(s\right)
g^{(q)}\left(s\varphi\left(\omega\right)\right)
\zeta^q2^{-\frac {n\left(q-1\right)}2}
\nonumber\\
 &=&
\frac1\zeta i-\sum_{q=1}^\infty
\lim_{n\rightarrow\infty}
\sum_{s=1}^{n}
 P_n\left(s\right)
g^{(q)}\left(s\varphi\left(\omega\right)\right)
\zeta^q2^{-\frac {n\left(q-1\right)}2}
\nonumber\\
&=&\frac1\zeta i-\frac \zeta{4\sin^2\frac\omega2}i
\end{eqnarray}
\vspace{-1cm}
\begin{equation}
\nonumber
\makebox[3cm]{} \zeta\in\mathbb{C},\quad \zeta\neq0,
\end{equation}
is derived by using  relation
(\ref{eq:asymptotics_of_alpha}) and lemma
(\ref{eq:limit_relation}). In the first equality, we make the Laurent
expansion at $\zeta=0$.  In the second equality, we just exchange the order
of the limit operations. In the last equality, we use relation
(\ref{eq:asymptotics_of_alpha}) and lemma (\ref{eq:limit_relation}).
Then, relation (\ref{eq:inportant_eignevalue}) is proved.

Next, we show the relations
\begin{eqnarray}
\lim_{n\rightarrow\infty} 
 \left<0_z|\psi_{\gamma_\pm}\right>\left<\psi_{\gamma_\pm}|0_x\right>
&=&\pm \frac{\exp\left(-\frac\omega2i\right)}2,
\label{eq:asymptotic_propaty_1}
\end{eqnarray}
\begin{equation}
\lim_{n\rightarrow \infty}\sum_{\gamma\neq\gamma_\pm}\left| \left<0_x|\psi_{\gamma}\right>\right|^2
=
\lim_{n\rightarrow \infty}\sum_{\gamma\neq\gamma_\pm}\left| \left<0_z|\psi_{\gamma}\right>\right|^2
=\:0,
\label{eq:asymptotic_propaty_2}
\end{equation}
where $\sum_{\gamma\neq\gamma_\pm}$ means the summation with respect to
all values $\gamma$ corresponding to eigenvalues of
$\hat G'\!\cdot\!\hat O'$ except for $\gamma_\pm$.  From relation
(\ref{eq:definition_of_eigenvector}),
\begin{eqnarray}
&&\lim_{n\rightarrow\infty}\left|\left<0_z|\psi_{\gamma_\pm}\right>\right|^{-2}
\nonumber\\
&=&
\lim_{n\rightarrow\infty}
\sum_{s=0}^{n}
\frac{P_n\left(s\right)\left|1-\exp\left(\omega i\right)\right|^2}
     {
\left|1-\exp\left(\gamma_\pm+s\varphi\left(\omega\right) \right)i\right|^2
}
\nonumber\\
&=&
1+\lim_{n\rightarrow\infty}
\sum_{s=1}^{n}
\frac{P_n\left(s\right)\left|1-\exp\left(\omega i\right)\right|^2}
     {
\left|1-\exp\left(\gamma_\pm+s\varphi\left(\omega\right) \right)i\right|^2
}
\nonumber\\
&=&2.
\label{eq:asymptotic_propaty_3}
\end{eqnarray}
In the second equality, we use
(\ref{eq:inportant_eignevalue}), and in the third equality, we use
(\ref{eq:inportant_eignevalue}), (\ref{eq:asymptotics_of_alpha}) and
(\ref{eq:limit_relation}).  On the other hand, from relation
(\ref{eq:original_relation}),
\begin{eqnarray}
\lim_{n\rightarrow\infty}
\frac{\left<0_z|\psi_{\gamma_\pm}\right>}
{\left<0_x|\psi_{\gamma_\pm}\right>}
&=&
\lim_{n\rightarrow\infty}\frac{1- \exp\left(\gamma_\pm i\right)}
{\left<0_x|0_z\right>\left(1-\exp\left(\omega i\right)\right)}
\nonumber\\
&=&\pm \exp\left(-\frac\omega2i\right)
\label{eq:asymptotic_propaty_4}
\end{eqnarray}
is derived. Using relations (\ref{eq:asymptotic_propaty_3})
and (\ref{eq:asymptotic_propaty_4}), relation
(\ref{eq:asymptotic_propaty_1}) is proved.  Furthermore, from 
(\ref{eq:asymptotic_propaty_3})
and (\ref{eq:asymptotic_propaty_4})
 and the trivial relation
\begin{eqnarray}
\sum_\gamma \left| \left<0_x|\psi_{\gamma}\right>\right|^2
\:=\:\sum_\gamma \left| \left<0_z|\psi_{\gamma}\right>\right|^2
\:=\:1,
\end{eqnarray}
(\ref{eq:asymptotic_propaty_2}) is derived.

Using some relations proved above, we obtain
\begin{eqnarray}
&&\lim_{n\rightarrow\infty} 
\left|\left<0_z\right|\left(\hat G'\!\cdot\!\hat O'\right)^{N'}\left|0_x\right>\right|
\nonumber\\
&=&
\lim_{n\rightarrow\infty} 
\left|\sum_{\gamma}\exp\left(N' \gamma i\right)
\left<0_z|\psi_\gamma\right>
\left<\psi_\gamma|0_x\right>
\right|\nonumber\\
&=&
\lim_{n\rightarrow\infty} 
\left|\exp\left(N'\gamma_+i\right)
\left<0_z|\psi_{\gamma_+}\right>
\left<\psi_{\gamma_+}|0_x\right>
\right.\nonumber\\&&{}\left.
+
\exp\left(N' \gamma_-i\right)
\left<0_z|\psi_{\gamma_-}\right>
\left<\psi_{\gamma_-}|0_x\right>
\right|\nonumber\\
&=&
\frac12\lim_{n\rightarrow\infty} 
\left|\exp\left(N' \gamma_+i\right)
-
\exp\left(N' \gamma_- i\right)
\right|\nonumber\\
&=&1.
\label{eq:end_of_proof_of_go}
\end{eqnarray}
In the second equality, we use relation
(\ref{eq:asymptotic_propaty_2}), in the third equality we use 
relation (\ref{eq:asymptotic_propaty_1}), and in the last one we use
(\ref{eq:inportant_eignevalue}) and (\ref{eq:suggestion_2}).  
Relation (\ref{eq:end_of_proof_of_go}) is exactly the same as
(\ref{eq:suggestion_1}). $\square$

\subsection{The case of more than one solution}
When there are two solutions, we also modify Grover's algorithm
in the same way. However,
 we have to know humming
distance $d$ of the two solutions. This information is not used in 
Grover's algorithm. 
When we change the number of solutions, all we have to do is to change
the definition of 
 $\varphi\left(\omega\right)$
and $N'$ as follows:
\begin{eqnarray}
\cot\frac\omega2&=&\sum_{s_1=0}^{n-d}\sum_{s_2=\delta\left(s_1,0\right)}^{d}\left(1+\left(-\right)^{s_2}\right)P_{n-d}\left(s_1\right)P_d\left(s_2\right)
\nonumber\\&&\makebox[2cm]{}\times\cot\frac{\left(s_1+s_2\right)\varphi_2\left(\omega\right)}2,
\nonumber
\end{eqnarray}
\vspace{-.5cm}
\begin{equation}
\makebox[.5cm]{}-\frac{2\pi}{n}<\varphi_2\left(\omega\right)<\frac{2\pi}{n},\quad 
\operatorname{sgn}\left(\omega\right)=\operatorname{sgn}\left(\varphi_2\left(\omega\right)\right),
\label{eq:definition_alpha_2_second}
\end{equation}
\begin{eqnarray}
N'_2&:= &\left[\frac{\pi}{4\sqrt2 \sin\frac\omega2}2^{\frac n2}+\frac12\right],
\label{eq:suggestion_2_second}
\end{eqnarray}
where the subscript ``$2$'' of $\varphi\left(\omega\right)$ and $N'$ indicates just the number of solutions.
Then, the operator $\hat G'$ and the oracle $\hat O'$ become
\begin{eqnarray}
\hat G'_2&:=&\exp\left(\varphi_2\left(\omega\right) \sum_{\alpha=0}^{n-1}S_x^{(\alpha)} i\right),\\
\hat O'_2&:=&
\exp\left(\omega
\left( \left|j_1\right>\left<j_1\right|+ \left|j_2\right>\left<j_2\right|\right)
 i\right).
\end{eqnarray}
The success probability goes to $1$ in the limit $n\rightarrow\infty$.
This is equivalent to the following relation:
\begin{eqnarray}
\lim_{n\rightarrow\infty}
\sum_{\eta=1,2}\left|\left<j_\eta\right|\left(\hat G'_2\!\cdot\!\hat O'_2\right)^{N'_2}\left|\bar 0\right>\right|^2
&=&1.
\label{eq:suggestion_1_second}
\end{eqnarray}
We can prove this relation in the same way as we have done in the one
solution case, so we omit it. We believe that the same
relations hold when there are more than two solutions, and we
numerically checked this fact in several cases.

\section{numerical calculation}
\label{sec:numerical_calculation}
\begin{table*}
\begin{tabular}{|c||r|r|r|r|r|r|}
\hline
   \# of items, i.e., $2^n$    &Grover &$\omega=\frac\pi2$  &$\omega=\frac{2\pi}3$&$\omega=\frac{3\pi}4$&$\omega=\frac{4\pi}5$&$\omega=1$
\\
\hline
\hline
$2^{10}$ &25                  &36                 &29 &27&26&25 \\
       &$5.4\times 10^{-4}$ &$2.2\times 10^{-1}$&$2.5\times 10^{-1}$ &$2.7\times 10^{-1}$ &$2.9\times 10^{-1}$&$6.8\times 10^{-1}$\\
\hline
$2^{20}$ &804                 &1137               &929                 &871 &846&804 \\
       &$2.4\times 10^{-7}$ &$8.5\times 10^{-2}$&$9.7\times 10^{-2}$ &$1.1\times 10^{-1}$ &$1.1\times 10^{-1}$&$6.2\times 10^{-1}$\\
\hline
$2^{30}$ &25735               &36396              &29717 &27856 &  27060&25736\\
       &$6.8\times 10^{-10}$&$5.0\times 10^{-2}$&$5.8\times 10^{-2}$ &$6.3\times 10^{-2}$ &$6.8\times 10^{-2}$&$6.1\times 10^{-1}$\\
\hline
$2^{40}$ &823549              &1164675            &950953              &891404 &865931&823550\\
       &$9.8\times 10^{-14}$&$3.5\times 10^{-2}$&$4.1\times 10^{-2}$ &$4.5\times 10^{-2}$ &$4.9\times 10^{-2}$&$6.0\times 10^{-1}$\\ 
\hline
\end{tabular}   
\caption{\label{fig:wide}
The upper integer in each cell indicates the optimal iteration number,
i.e., $N$ or $N'$.  The lower real number in each cell indicates the error
rate, i.e.,
$1-\left|\left<j\right|\left(\hat G\!\cdot\!\hat O\right)^N\left|\bar0\right>\right|^2$
 or
$1-\left|\left<j\right|\left(\hat G'\!\cdot\! \hat O'\right)^N\left|\bar0\right>\right|^2$.
The optimal iteration number and error rate in the case of Grover's algorithm
are in the leftmost column, and those for the algorithm
using the proposed  quantum circuit at
$\omega=\frac12\pi,\frac23\pi,\frac34\pi,\frac45\pi,\pi$ are in
the other columns.
Note that, $N'$ and $\varphi\left(\omega\right)$ when $\omega=\pi$
are defined in the same way as the other four examples, i.e. (\ref{eq:suggestion_2}) and
 (\ref{eq:definition_alpha_2}).
 However, as pointed out in sec. \ref{sec:G_algorithm}, relation (\ref{eq:suggestion_1}) does not hold in that case.
 All the values were calculated for the case of only one
solution. 
}
\end{table*}
In order to check that the new  quantum circuit works well, we numerically
calculated the iteration number, i.e., $N'$ defined by
(\ref{eq:suggestion_2}), and the error rate, i.e.,
$1-\left|\left<j\right|\left(\hat G'\!\cdot\!\hat O'\right)^{N'}\left|\bar0\right>\right|^2$,
at $n=10,20,30,40$ and $\omega=\frac\pi2,\frac{2\pi}3,\frac{3\pi}4,\frac{4\pi}5,\pi$.  The results are
shown in Table \ref{fig:wide}.  In order to compare the
proposed  quantum circuit with the  quantum circuit used in Grover's algorithm, we
also show the corresponding values for Grover's algorithm in the
 table.
Note that, $N'$ and $\varphi\left(\omega\right)$ when $\omega=\pi$
are defined in the same way as the other four example, i.e. (\ref{eq:suggestion_2}) and
 (\ref{eq:definition_alpha_2}).

From the result when $\omega=\pi$,
we predict that the relation
\begin{eqnarray}
\lim_{n_o\rightarrow\infty}\inf_{n_0<n} 
\left|\left<j\right|\left(\hat G'\!\cdot\!\hat O'\right)^{N'}\left|\bar 0\right>\right|^2
&=&Const
\nonumber\\
0<Const<1\makebox[-1cm]{}
\end{eqnarray}
holds when case $\omega=\pi$. This relation may be proved in a way
similar to that in the other $\omega$ case.
This relation means that we can probably use the quantum circuit, i.e.
 $\left(\hat G'\!\cdot\!\hat O'\right)^{N'}$, for the quantum search problem even when $\omega=\pi$,
though the error rate for the circuit will be much bigger than that in other
$\omega$ cases.

What we want to mention about the results for the cases $\omega=\frac\pi2,\frac{2\pi}3,\frac{3\pi}4,\frac{4\pi}5$ is that the error rate is
sufficiently small for realistic $n$ cases. 
 On the other
hand, it is fair to point out that with the algorithm using the new
quantum circuit, the number of iterations
 and the error rate are much higher than  in Grover's
algorithm.  However,   the results do not provide  enough 
information for us to
discuss the efficiency of the two algorithms.  We remark that operator
$\hat G$ is a really multi-particle operator, whereas
operator $\hat G'$ is just a set of single-particle rotation, i.e.,
a direct product of single-qubit operators.
The  ``really multi-particle operators'' are those that
 can not be expressed only by
products of single-qubit operators.
 Therefore, operator $\hat G'$
can be executed much faster than $\hat G$ in many realistic
systems. Then, the average time to find solution $j$ by the algorithm using
the new  quantum circuit is shorter than that by Grover's algorithm in some
cases on a realistic QC.

\section{relation between the proposed  quantum circuit and the AQC}
\label{sec:relation_AQC} The  quantum circuit proposed in this paper is inspired
by Farhi's Hamiltonian \cite{FGG+00} for the AQC.  In this section, we
briefly  review the AQC, point out the simple relation between the
 quantum circuit used in Grover's algorithm and Roland's
Hamiltonian \cite{RC01} for quantum search on the AQC, and finally point out
the similar relation between the proposed  quantum circuit and Farhi's
Hamiltonian for quantum search on the AQC.  Recall that
to generate a new  quantum circuit, we assumed the existence of operator $\hat G'$ related to
 Farhi's Hamiltonian as an analogy of the relation between the Grover
operator and Roland's Hamiltonian. This relation is shown
below. Then, we find the explicit expression of operator $\hat G'$,
i.e., (\ref{eq:def_oracle_Go_G'}).

The AQC involve the  following procedures.  First, we
define the parametrised hermitian matrix $\hat H\left(r\right)$ that
has the following five properties.
\begin{itemize}
 \item The operator $\hat H\left(r\right)$ is continuously changed with
respect to parameter  $r\in\mathbb{R}$.
 \item  The ground state of $\hat H\left(0\right)$ is
a simple general state.
 \item The ground state of $\hat H\left(1\right)$ is
an encoded solution of the problem.
 \item At any $0\leq r\leq1$, the ground state of $\hat H\left(r\right)$
does not degenerate. 
\item The Hamiltonian can be easily defined using only the definition of
the problem, i.e.,  the Hamiltonian  can be defined without knowing the result.
\end{itemize}
Second, we prepare the initial state that is the ground state of $\hat
H\left(0\right)$.  Third, we make the time evolution of the state such
that
\begin{eqnarray}
i\frac\partial{\partial t}\left|\phi_T\left(t\right)\right>
&=&
\hat H\left(\mu\left(\frac tT\right)\right)\left|\phi_T\left(t\right)\right>
\nonumber
\end{eqnarray}
\vspace{-.6cm}
\begin{equation}
\makebox[2.1cm]{}
\mu\left(0\right)=0\quad
\mu\left(1\right)=1\quad
\frac d{dr}\mu\left(r\right)>0.
\end{equation}
Note that $\mu\left(r\right)$ can be chosen arbitrarily until the above
conditions are satisfied, but the choice affects the probability of
success and the time for the calculation.  Finally, we observe the state
at time $t=T$.  If $T$ is sufficiently large, the correct
solution is obtained, i.e.,
\begin{eqnarray}
\lim_{T\rightarrow\infty}
\left| \left<
 \phi_g\left(r\right)|\phi_T\left(r T\right)
\right>\right|
&=&1
\end{eqnarray}
where $\left|\phi_g\left(r\right)\right>$ is a ground state of the
operator $\hat H\left(r\right)$. A suitable value of $T$ can be found
from the adiabatic theorem.  This is a rough sketch of the AQC.

Next, we show the relation between the  quantum circuit used in Grover's
algorithm and Roland's Hamiltonian for quantum search \cite{RC01} on the
AQC.  The
Hamiltonian
\begin{eqnarray}
\hat H_R\left(r\right)
&:=&
-\left(1-r\right)\left|\bar0\left>\right<\bar0\right|-r\left|j\right>\left<j\right|
\label{eq:AQC_QC_1}
\label{eq:hamiltonian_roland}\\
\mu_R\left(r\right)
&:=&
 \frac{\sin\left(\pi-2\theta\right)r}{\sin\left(\pi-2\theta\right)r+\sin\left(\left(\pi-2\theta\right)r+2\theta\right)}
\label{eq:AQC_QC_2}
\end{eqnarray}
executes quantum search, where $\left|\bar0\right>$ and $\theta$ mean
the same state and value as those in the previous section,
i.e., (\ref{def:initial_state}) and (\ref{def:theta}), and $j$ is the
target of the search.  The above function $\mu_R\left(r\right)$ is optimised so as
to maximize the success probability.  From this expression, it is readily
known that
\begin{eqnarray}
\hat G&=&\exp\left(i\pi2\left(1-\mu_R^*\right)\hat H\left(0\right)\right)\nonumber\\
\hat O&=&\exp\left(i\pi2        \mu_R^*       \hat H\left(1\right)\right), 
\label{rel:Grover_adiabatic_1}
\end{eqnarray}
where the operators $\hat G$ and $\hat O$ are defined by
(\ref{eq:def_oracle_Gr}) and $\mu_R^*$ satisfies the condition that
 the gap between the two lowest eigenvalues of 
$\hat H_R\left(r\right)$ becomes the minimum value  at the point
$r=\mu_R^*$.  Furthermore, by some calculations, we
can check that
\begin{eqnarray}
\lim_{T\rightarrow\infty}\left|\left<\phi_T\left(\frac{4\theta T}{\pi-2\theta}m\right)\right|\left(\hat G\!\cdot\!\hat O\right)^{m}\left|\bar0\right>\right|
&=&1
\label{rel:Grover_adiabatic_2}
\end{eqnarray}
where $0\leq m\leq\left[\frac\pi{4\theta}+\frac14\right]$ is an integer.
This relation means that the optimal speed of an AQC using Roland's
Hamiltonian is exactly the same as the speed of Grover's algorithm
with respect to quantum search.

Next, we show the relation between the  quantum circuit proposed in this paper
and Farhi's Hamiltonian \cite{FGG+00} for quantum search on an AQC.  The Hamiltonian
\begin{eqnarray}
 \hat H_F\left(r\right)
&:=&
-\left(1-r\right)\sum_{\alpha=0}^{n-1}S^{(\alpha)}_x-r\left|j\right>\left<j\right|
\label{eq:hamiltonian_farhi}
\end{eqnarray}
also executes quantum search.  As is easily shown, the following
relation holds
\begin{eqnarray}
\hat G'&=&\exp\left(i\pi\xi\left(1-\mu_F^*\right)\hat H_F\left(0\right)\right)
\nonumber\\
\hat O'&=&\exp\left(i\pi\xi        \mu_F^*       \hat H_F\left(1\right)\right),
\label{rel:Go_adiabatic_1}
\end{eqnarray}
where $\xi:=\omega/\mu_F^*$ is a real number. Relations
(\ref{rel:Grover_adiabatic_1}) and (\ref{rel:Go_adiabatic_1}) are very
similar. 
However, we can only check that the leading term
of $\mu_F^*$ as a function of $n$ is the same as that of $\mu_F^{*\prime}$,
 where at the point
$r=\mu_F^{*'}$ the gap between the two lowest eigenvalues of 
$\hat H_F\left(r\right)$ becomes the minimum value. 
  Unfortunately, we have not yet found
a relation like (\ref{rel:Grover_adiabatic_2}) in this case.

What we want to say in this section is that there are some relations
between  the quantum circuits for the QC and the Hamiltonians for the AQC, and
these relations can be used to generate  new  quantum circuits.  Some people may
think that these relations are trivial or just accidental things.
However, it is a truth that the proposed  quantum circuit is found on
the basis of  the
conviction that there must be an operator $\hat G'$ related to
(\ref{eq:hamiltonian_farhi}) as an analogy of the relation between $\hat
G$ and (\ref{eq:hamiltonian_roland}),
i.e., (\ref{rel:Grover_adiabatic_1}) and (\ref{rel:Grover_adiabatic_2}).
Accordingly, we believe that there are more hidden relations between
 quantum circuits and Hamiltonians and that they would be powerful instruments for
generating new  quantum circuits and new Hamiltonians.

\section{conclusion}
We have proposed a new quantum circuit for the quantum search problem.
This quantum circuit is superior to the quantum circuit used in
Grover's algorithm in some cases on a realistic quantum computer. The
reasons for this superiority in short are as follows: In the
quantum circuit proposed in this paper, all the operators except for the
oracle are direct products of single-qubit gates.  In the quantum
circuit used in Grover's algorithm, there are the operators other than
 the oracle, which are really multi-particle operators.  On the other
hand, it is a fact that the product of single-qubit gates can be
executed much faster than multi-particle operators in many
realistic systems.
In addition, the scaling of the number of oracle calls for this circuit
is the same as that for Grover's algorithm,
i.e. $O\left(2^{n/2}\right)$.

The proposed circuit is found by  a comparison
of circuits for the quantum computer and Hamiltonian for the adiabatic quantum computer.
This fact indicates that the comparison is probably one of the powerful instruments for
finding efficient new quantum circuits.

One aspect of future work is to find a
stricter relation between the quantum circuits for the quantum
computer and the Hamiltonians for the adiabatic quantum computer that gives
sufficient
data for modification from the Hamiltonians into the
 quantum circuits.
Then,
we will be able to  automatically generate other efficient quantum
circuits from Hamiltonians for the adiabatic quantum computer with respect
to other problems that the adiabatic quantum computer is good at and
discover new concepts for
quantum circuits.

\section*{acknowledgements}
The author wish to thank Y. Kawano, S. Tani, Y. Takahashi and
Y. Nakajima for discussions and valuable comments.

\appendix
\section{Proof of Lemma (\ref{eq:limit_relation})}
\label{sec:proof_of_lemma}
Here, we prove lemma (\ref{eq:limit_relation}).

{\it Proof}:

The sufficient condition of (\ref{eq:limit_relation}) is the relation
\begin{equation}
 \lim_{n\rightarrow\infty}
\sum_{s=1}^{n}P_n\left(s\right)\left(\frac sn\right)^q
=2^{-q}\nonumber
\label{eq:sufficient_condition}
\end{equation}
\begin{equation}
q\in\mathbb{Z}, 
\end{equation}
We can check this as follows:
\begin{eqnarray}
&& \lim_{n\rightarrow\infty}\sum_{s=1}^{n}P_n\left(s\right)f\left(\frac sn\right)\nonumber\\
&=&
 \lim_{n\rightarrow\infty}\sum_{s=1}^{n}P_n\left(s\right)
\left(
\sum_{q=-\alpha}^{-1}\left(\frac sn\right)^qf^{(q)}
\right.\nonumber\\&&\left.{}
\makebox[2cm]{}+\sum_{q=0}^{\infty}\left(\frac sn-\frac12\right)^qf^{(q)}
\right)
\nonumber\\
&=&
\sum_{q=-\lambda}^{-1}f^{(q)}
 \lim_{n\rightarrow\infty}\sum_{s=1}^{n}P_n\left(s\right)
\left(\frac sn\right)^q
\nonumber\\&&{}
{}+\sum_{q=0}^{\infty}f^{(q)}
 \lim_{n\rightarrow\infty}\sum_{s=1}^{n}P_n\left(s\right)\left(\frac sn-\frac12\right)^q
\nonumber\\
&=&
\sum_{q=-\lambda}^{-1}f^{(q)}
2^{-q}+f^{(0)}
\nonumber\\
&=&f\left(\frac12\right),
\end{eqnarray}
where $f^{(q)}$ is defined as
\begin{eqnarray}
 f\left(\zeta\right)=
\sum_{q=-\lambda}^{-1}\zeta^qf^{(q)}
+\sum_{q=0}^{\infty}\left(\zeta-\frac12\right)^qf^{(q)}.
\end{eqnarray}
The (\ref{eq:sufficient_condition}) is used in the third equality. The other
equalities are easily given from the above definition of $f^{(q)}$

In the rest of this appendix, we prove relation
(\ref{eq:sufficient_condition}).  We define some functions,
\begin{eqnarray}
 F\left(n,q\right)&:=&\sum_{s=1}^{n}P_n\left(s\right)\left(\frac sn\right)^q\\
n^q \tilde F\left(n,q\right)&:=&\sum_{s=\max\left(q,1\right)}^{n}\frac{s!}{\left(s-q\right)!}P_n\left(s\right),
\\
&=&
\left\{\begin{array}{l }
\frac{n!2^{-q}}{\left(n-q\right)!}
\makebox[1cm]{} \makebox{ in case of }q>0\\
\frac{n!2^{-q}}{\left(n-q\right)!}-
\sum_{s=q}^{0}\frac{n!2^{-n}}{\left(s-q\right)!(n-s)!}
\\
\makebox[2cm]{} \makebox{ in case of }q\leq 0.
       \end{array}\right.
\end{eqnarray}
From these definitions, we can derive the relation
\begin{eqnarray}
&&\tilde F\left(n,q\right) \nonumber\\
&\leq& F\left(n,q\right)   \nonumber\\
&\leq& \left(\frac{n}{n-4q}\right)^q\tilde F\left(n,q\right)+\sum_{s=1}^{\left[\frac n4\right]+1}P_n\left(s\right)\left(\frac sn\right)^q\!\!\!.
\end{eqnarray}
for $n>4q$.  Using the following relation
\begin{eqnarray}
\lim_{n\rightarrow \infty} n!\frac{e^n}{n^{n+\frac12}\sqrt{2\pi}}&=&1,
\end{eqnarray}
we can see that both the upper bound and the lower bound of
$F\left(n,q\right)$ goes to $2^{-q}$ in the limit $n\rightarrow\infty$. $\square$

\end{document}